\documentclass[12pt]{article}
\usepackage{amsmath}
\usepackage{amssymb}
\usepackage{amsthm}
\usepackage{amsfonts}
\usepackage{graphicx}
\usepackage{color}

\numberwithin{equation}{section}

\newcommand{\R}{{\mathbb R}}
\newcommand{\Z}{{\mathbb Z}}

\newcommand{\wt}[1]{\widetilde{#1}}
\newcommand{\wh}[1]{\widehat{#1}}
\newcommand{\vv}[1]{\marginpar{\small\sf #1}}
\newcommand{\ed}{\mathrm{d}}
\newcommand{\pr}{\textbf{{Proof }}}
\newcommand{\reff}[1]{(\ref{#1})}

\newtheorem{theorem}{Theorem}[section]
\newtheorem{lemma}[theorem]{Lemma}

\newtheorem{corollary}[theorem]{Corollary}

\theoremstyle{definition}

\title{ Percolation properties of the non-ideal gas }

\author{
E.~Pechersky${}^\dag$ and A.~Yambartsev${}^\ddag$\\
\footnotesize{\noindent${}^\dag$ IITP, 19, Bolshoj Karetny per.,
Moscow, Russia}
\\  \footnotesize{pech@iitp.ru}\\
\footnotesize{\noindent${}^\ddag$ IME-USP, Rua do Mat\~{a}o, 1010,
05508-090, S\~{a}o
Paulo, Brazil} \\
\footnotesize{yambar@ime.usp.br}}

\date{}
\begin{document}
\maketitle

\begin{abstract} We estimate locations of the regions of
the percolation and of the non-percolation in the plane
$(\lambda,\beta)$: the Poisson rate -- the inverse temperature,
for interacted particle systems in  finite dimension Euclidean
spaces. Our results about the percolation and about the
non-percolation are obtained under different assumptions. The
intersection of two groups of the assumptions reduces  the results
to two dimension Euclidean space, $\R^2$, and to a potential
function of the interactions having a hard core.

The technics for the percolation proof  is based on a contour
method which is applied to a discretization  of the Euclidean
space. The technics for the non-percolation proof is based on the
coupling of the Gibbs field with a branching process.



\end{abstract}

\section{Introduction}
A rigorous proof of phase transitions for continuous models of the
statistical mechanics is still an open problem if the interactions
between particles are described by conventional in physics
potential functions. The first and yet to this moment the only
example of the rigorous proof of the phase transition in a
continuous model is the result by J.L.~Lebowitz, A.~Mazel and
E.~Presutti in \cite{LMP}. The potential functions in \cite{LMP}
are a pre-limiting version of the mean-field interaction, see
\cite{KUH}, having a large but finite radius of the interactions,
and a four-body stabilizing potential function.

In the present work, we investigate the percolation properties of
an interacting particle ensemble. We describe a phase diagram of
the continuous system in the plane $(\lambda,\beta)$, \textit{the
Poisson rate -- the inverse temperature}. The interaction is
defined by pair potential functions. We do not prove the phase
transition driven by boundary conditions  as in \cite{LMP}.
However we think that the transition: \textit{the percolation --
the non-percolation} can be considered as a phase transition
relatively, for example, to the conductivity of the matter or the
velocity of the sound propagations.

The book \cite{MS} gives a rather complete picture of the state of
the continuum percolation theory for the ideal gas from the
mathematical point of view. Much attention in \cite{MS} is drawn
to  the Boolean percolation problem for the Poisson point
processes in $\R^\nu$. Points of a configuration of the process
are considered as the centers of closed balls  of a random radius
(\textit{Boolean radius}) such that the radii corresponding to
different points are independent of each other (and also
independent of the process) and identically distributed. The
existence of an unbounded connected component in the set composed
by the union of all random balls means the percolation. The
unbounded component is called an infinite cluster. One of the main
results in \cite{MS} which is related to the present article is
about the existence of a critical value $\lambda_c$ of the rate of
the Poisson point processes. Namely, the value $\lambda_c$
distinguishes the percolation and the non-percolation, where the
last means that with the probability 1 only bounded connected
components exist in the union of the balls. It is asserted in
\cite{MS} that with the probability 1 there are no infinite
clusters when $\lambda<\lambda_c$ and there exists an infinite
cluster when $\lambda>\lambda_c$.

We consider the same problem but for  a non-ideal gas, which is
determined by some interaction potential  function and the Poisson
free measure with the rate $\lambda.$ A non-percolation condition
was studied in the work \cite{Mur} for  positive finite range
potential functions. The Boolean radius is equal to the range of
the interaction. A new proof of this result can be found in
\cite{zes}.

Next we give a brief description of our results  not concerning
the conditions. We study the case when the potential function
takes as positive as negative values. The potential function
determines a Gibbs measure of which the percolation properties we
investigate. In this case a new parameter enters into the game it
is temperature $T$. By a tradition we more often use the inverse
temperature $\beta=\frac{1}{T}$. The results we present here
outline two regions in the plane $(\lambda,\beta)$ of the
percolation and the non-percolation with probability 1 for a
Boolean radius $\ell$. We do not seek the solution as precise as
possible. Our aim is to outline the regions such that they have
typical forms. Namely (see Figure \ref{fig1}):

The region of the non-percolation can be described as follows.

\vspace{.5cm}

\textit{There exists a density value $\lambda^-_\ell$ such that
for any $\lambda<\lambda^-_\ell$ there exists an inverse
temperature $\beta^-_\ell(\lambda)$ such that for any
$\beta<\beta^-_\ell(\lambda)$ all clusters are finite with Gibbs
probability 1. }

\vspace{.5cm}

The region of the percolation can be described as follows.

\vspace{.5cm}

\textit{For any density $\lambda$ there exists an inverse
temperature $\beta^+_\ell(\lambda)$ such that for all
$\beta>\beta^+_\ell(\lambda)$ there exists an infinite cluster
with Gibbs probability 1. There exists a density value
$\lambda^+_\ell$ such that $\beta^+_\ell(\lambda)=0$ if
$\lambda>\lambda^+_\ell$}

\vspace{.5cm}

Our results provide estimates of the parameter regions separating
the areas of the existence $A_+$ and of the non-existence $A_-$ of
an infinite cluster. There exists a region between $A_+$ and $A_-$
where our result does not give the answer on the percolation.

The result shows in particular that for any small density
$\lambda$ there exists an infinite cluster if temperature is low
enough. Another feature is also that the non-existence of an
infinite cluster may only be  at a small density,
$\lambda<\lambda^-_\ell$. This fact is in accord with the result
(\cite{MS}) of the Boolean non-percolation for the ideal gas.

The conditions under which we prove the percolation and  the
non-percolation are different. We prove the percolation result in
$\R^2$ only.  It is necessary as well, that the potential function
has an attractive part. However, we do not assume a hard core. Our
proof of the non-percolation requires the hard-core condition. The
attractive part of the potential as well as the dimension of the
Euclidean space are not restrictions for our proof of the
non-percolation . Besides, all results are proved for a non-random
Boolean radius $\ell$.

The technics for the proofs of the existence and of the
non-existence of an infinite cluster drastically differ. For the
existence of an infinite cluster we use technics close to the
contour methods (see \cite{Fernandez}).  The non-existence is
proved by a coupling of the Gibbs state and a branching process.
The extinction of the branching process leads to the non-existence
of infinite clusters. We use a branching process with interactions
between offsprings in the same generations and between the
generations. The hard core condition prevents the accumulations of
a large offspring amount which can appear because of attractive
interactions between the offsprings.

The section \ref{Mod} contains the definitions, all assumptions
and the formulations of the main results. All proofs are in the
section \ref{pro}.

\section{Model and Results}\label{Mod}
\subsubsection*{The configuration space, the potential function and Hamiltonian}
The non-ideal gas model is a pair $(\Omega,\varphi)$. Here
$\Omega=\{\omega\}$ is the set of all countable subsets in
${\mathbb R}^\nu$ such that for any bounded $V\subset {\mathbb
R}^\nu$
\begin{equation}\label{0}
\#(\omega\cap V)<\infty,
\end{equation}
where $\#(W)$ is the number of points in $W$. $\omega$ is the set
of points from ${\mathbb R}^\nu$, where particles $x\in\omega$
sit. The set $\Omega$ is called the set of
\textit{configurations}. We use the standard notations for the
restrictions on subsets. If $V$ is a Borel set in ${\mathbb
R}^\nu$ and $\omega\in\Omega$ then $\omega_{V}=\omega\cap V$ and
$\Omega_{V}$ is the set of all configurations in $V$. If $V\cap
V'= \varnothing$ and $\omega\in\Omega$ then $\omega_{V\cup V'}=:
\omega_{V} \vee \omega_{V'}$.

The  $\sigma$-algebra $\mathfrak{A}$ in $\Omega$ is generated by
the cylinder sets
\begin{equation}\label{ev}
\mathcal{A}_{V,n}:= \{\omega: \:\#( \omega_{V} )= n\} \subseteq
\Omega,
\end{equation}
where $V$ is a bounded Borel set in $\mathbb R^\nu.$

The potential function $\varphi$ describes the interaction of the
particles. We consider pair interactions only and  assume that
$\varphi(x,y)$ is continuous and satisfies the following
properties.
\begin{itemize}
\item Translation invariance: for any $(x,y)\in {\mathbb
R}^\nu\times {\mathbb R}^\nu$ and any $z\in {\mathbb R}^\nu$ it
holds that $ {\varphi} (x+z,y+z)={\varphi} (x,y). $

Therefore we can introduce the function
$\wh{\varphi}(x),\;x{\in\mathbb R}^\nu$,  by the equality $
\wh{\varphi}(x-y)={\varphi}(x,y), $ which further we denote with
the same symbol $\varphi(x)$.

\item Isotropy: if $B$ is an orthogonal operator in $\R^\nu$ then
$\varphi(Bx)=\varphi(x)$.

\item There are two reals $f\geq 0$, $d>0$  such that  $f\leq d$
and
\begin{equation}\label{phi}
\varphi(x)\begin{cases}
=\infty,&\mbox{ if }|x|\leq f,\\
\geq 0,&\mbox{ if }|x|\in [f,d],\\
\leq 0,&\mbox{ if }|x|\in [d,\infty).
\end{cases}
\end{equation}
Besides there exists  a positive monotone decreasing function
$\psi$ and $g>d$ such that

\begin{align}
\varphi(x)\geq -\psi(x)\mbox{ for }x\geq g,\\
\intertext{and}
 \label{tail}I=\int_g^\infty r^{\nu-1}\psi(r)\ed r
<\infty.
\end{align}
The condition \reff{tail} was proposed in \cite{D}.

\item Lower boundedness: there exists $M>0$ and $x_0$ such that
$\min_x \varphi(x)= \varphi(x_0)=-M$.
\end{itemize}

Hamiltonian is
\begin{equation} \label{H} H(\omega) = \sum_{x,y\in \omega}
\varphi(x-y)
\end{equation}
which describes energy of configuration $\omega$. The above
expression is formal since the sum does not exists. The energy of
$\omega_{V}\in \Omega_{V}$ with boundary condition $\tau\in
\Omega_{V^{c}}$ is
\begin{equation}\label{condham}
H(\omega_{V} \mid \tau )=H(\omega_{V})+F(\omega_{V},
\tau):=\sum_{x,y\in \omega_{V}}\varphi(x-y)+ \sum_{x\in
\omega_{V},y\in \tau}\varphi(x-y).
\end{equation}
The last sum in the above expression might be infinite. However,
if $f>0$ and \reff{tail} holds then $\sum_{x\in \omega_{V},y\in
\tau}\varphi(x-y)<\infty$ for any finite $V$, and any $\omega_V$
and $\tau$.

\subsubsection*{The reference and the Gibbs measures}
The  reference measure $\Pi$ is defined as Poisson one on
$(\Omega,\mathfrak{A})$ with intensity $\lambda>0:$
\begin{equation}\label{refm}
\Pi(\mathcal{A}_{V,n}) = \frac{\lambda^n
|V|^n}{n!}e^{-\lambda|V|},
\end{equation}
where $|V|$ is the volume of $V$ (see \reff{ev}). The Gibbs
measure $P^{\beta, \lambda}$  on $(\Omega,\mathfrak{A})$ is
determined by the Gibbs reconstruction method of the reference
measure (see \cite{MM}).

To define $P^{\beta, \lambda}$ we introduce a Gibbs specification
$$
\{P^{\beta,\lambda}_{V,\tau},\  V\subset \mathbb R^\nu, \tau\in
\Omega_{V^{c}} \}
$$
which is a family of the Gibbs reconstruction of the measure $\Pi$
in finite volumes $V$ given a conditional configuration $\tau$ and
the inverse temperature $\beta\in \mathbb R_+.$ The measure
$P^{\beta,\lambda}_{V,\tau}$ has the following density
$p^{\beta,\lambda}_{V,\tau}$ with respect to the measure $\Pi:$
\begin{equation}\label {Gibbs}
p^{\beta,\lambda}_{V,\tau}(\omega_{V}) = \frac{\exp \{ -\beta
H(\omega_{V} \mid \tau) \}} {\int_{\Omega_{V}} \exp \{ -\beta
H(\omega \mid \tau) \}\Pi(\ed\omega_{V}) }.
\end{equation}
It is assumed that the densities
$p^{\beta,\lambda}_{V,\tau}(\omega_{V})$ are defined for boundary
configurations $\tau$ such that \reff{condham} is finite.

We assume some conditions for the existence of the integral in
\reff{Gibbs} and for the existence of at least one of Gibbs
measures $P^{\beta,\lambda}$ corresponding to the specification
\reff{Gibbs} (see \cite{Ru} or \cite{PZh}). If $H(\cdot|\tau)$ is
finite not for all $\tau$ then the existence conditions is such
that the Gibbs measure $P^{\beta,\lambda}$ is concentrated on a
set of the configurations, where $H(\cdot|\tau)$ is finite, when
$\tau$ from this set (see \cite{S}).

Further we use the notation $H_{V}(\cdot|\cdot)$ for the energy of
configurations from $\Omega_{V}$ with a boundary condition. We
shall often omit some indices and shall write $P_{V}$ instead of
$P^{\beta,\lambda}_{V,\tau}$ and $P$ instead of
$P^{\beta,\lambda}$.

\subsubsection*{The percolation} Any ordered sequence of particles from
a gas configuration $\omega$ we shall call a path. We say that two
points $x,y\in \R^\nu$ $\ell$-percolate with respect to $\omega$
if there exists some finite path $\pi = \{ x_1, x_2, \dots,
x_n\}\subset\omega$ such that $|x_i-x_{i-1}|\le \ell$ for all
$i=2,\dots,n,$ and $|x-x_{1}|\le \ell,\ |y-x_{n}|\le \ell$. The
set $\pi$ is called a $\ell$-\textit{cluster} in $\omega$ or
simply a \textit{cluster} when $\ell$ and $\omega$ are fixed. We
say that $x,y\in \R^\nu$ in a cluster $\pi$ if they
$\ell$-percolate with respect to $\omega$. We shall denote by
$(x\leadsto^\ell y)$ the event that $x$ and $y$ are in a cluster.
If there exists an infinite cluster starting at $x\in \R^\nu$,  we
denote this event by $(x\leadsto^\ell \infty).$ The probability of
the event $(0\leadsto^\ell \infty)$ we call {\it
$\ell$-percolation function} or  \textit{percolation function} and
denote $\theta_\ell(\beta,\lambda):$
\begin{equation}\label{pf}
\theta_\ell(\beta,\lambda):=P^{\beta,\lambda} (0\leadsto^\ell
\infty ).
\end{equation}

\subsubsection*{The main result on the non-percolation} The result
on the non-percolation is proved for $f>0$ (the hard-core
condition) and any $\nu$ (any dimension of the Euclidean space).

\begin{theorem}\label{hard-core} For any $\ell> f$ there
exists $\lambda^-_\ell,\ 0<\lambda^-_\ell<\infty$,  and a function
$\beta^-_\ell(\lambda) $ defined on the interval
$(0,\lambda^-_\ell)$ such that
\begin{enumerate}
    \item $0<\beta^-_\ell(\lambda)<\infty$ on the interval
    $(0,\lambda^-_\ell)$,
    \item $\beta^-_\ell(\lambda)\uparrow \infty$ as $\lambda\downarrow
    0$.
\end{enumerate}
Let
\begin{equation*}
A_-=\{(\lambda,\beta):\:\lambda<\lambda^-_\ell,\
\beta<\beta^-_\ell(\lambda)\}.
\end{equation*}
Then  all clusters are finite with the probability 1 if
$(\lambda,\beta)\in A_-$, that is $\theta_\ell(\beta, \lambda)=0$
for those $(\lambda,\beta)$ (see Fig.\ref{fig1}). Moreover the
expectation of the cluster size is finite.
\end{theorem}

\subsubsection*{The main result on the percolation} 
\begin{figure}
\begin{center}
\includegraphics[scale=0.3]{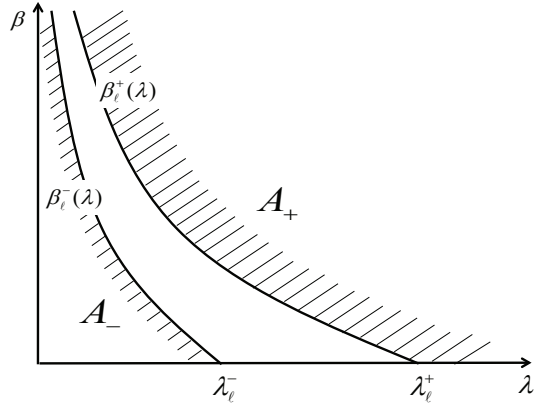}
\end{center}
\caption{Percolation and non-percolation regions.}%
\label{fig1}
\end{figure}

The next theorem is proved
for the case $\nu=2$. Now the Boolean radius is bounded below,
however the hard core is not necessary.
\begin{theorem}\label{t2}
For any $\ell>2\sqrt{2}d$ there exist $\lambda^+_\ell,\
0<\lambda^+_\ell<\infty$,  and a function $\beta^+_\ell(\lambda) $
defined on the interval $(0,\lambda^+_\ell)$ such that
\begin{enumerate}
    \item $0<\beta^+_\ell(\lambda)<\infty$ on the interval
    $(0,\lambda^+_\ell)$,
    \item $\beta^+_\ell(\lambda)\uparrow \infty$ as $\lambda\downarrow
    0$.
\end{enumerate}
Let
\begin{equation*}
A_+=\{(\lambda,\beta):\:\lambda<\lambda^+_\ell,
\beta>\beta^+_\ell(\lambda)\}\cup\{(\lambda,\beta):\:\lambda>\lambda^+_\ell,
\beta\geq 0\}.
\end{equation*}
Then with the probability 1 there exists an infinite cluster  if
$(\lambda,\beta)\in A_+$, that is $\theta_\ell(\beta, \lambda)>0$
for those $(\lambda,\beta)$ (see Fig.\ref{fig1}).
\end{theorem}

\vspace{1cm.}

The next corollary joins the results of theorems \ref{hard-core}
and \ref{t2}. However, the claims of the corollary hold under the
most restrictive assumptions from the assumptions of both
theorems.

\begin{corollary}
Let $\nu=2$ and $f>0$. For all $\ell>2\sqrt{2}d$ there exist two
positive reals $\lambda^-_\ell$ and $\lambda^+_\ell$, and two
functions: $\beta^-_\ell(\lambda)$ defined on
$(0,\lambda^-_\ell)$, and  $\beta^+_\ell(\lambda)$ defined on
$(0,\lambda^+_\ell)$, such that
\begin{enumerate}
    \item $0<\lambda^-_\ell<\lambda^+_\ell<\infty$,
    \item $0<\beta^-_\ell(\lambda)<\infty$ and
    $0<\beta^+_\ell(\lambda)<\infty$ on their intervals of
    the definitions,
    \item $\beta^-_\ell(\lambda)<\beta^+_\ell(\lambda)$ on
    $(0,\lambda^-_\ell)$,
    \item $\beta^-_\ell(\lambda)\uparrow \infty$ and
    $\beta^+_\ell(\lambda)\uparrow \infty$ as $\lambda\downarrow
    0$.
\end{enumerate}
Let
\begin{align*}
A_-&=\{(\lambda,\beta):\:\lambda<\lambda^-_\ell,\
\beta<\beta^-_\ell(\lambda)\}\\
\intertext{and}
 A_+&=\{(\lambda,\beta):\:\lambda<\lambda^+_\ell,\
\beta>\beta^+_\ell(\lambda)\}\cup\{(\lambda,\beta):\:\lambda>\lambda^+_\ell,\
\beta\geq 0\}.
\end{align*}
Then
\begin{enumerate}
    \item with the probability 1 all clusters are finite  if $(\lambda,\beta)\in
A_-$,
    \item with the probability 1 there exists an infinite cluster
    if $(\lambda,\beta)\in A_+$,
\end{enumerate}
(see Fig.\ref{fig1}).

Moreover the expectation of the cluster size is finite if
$(\lambda,\beta)\in A_-$.
\end{corollary}

\section{Proofs}\label{pro}
\subsection{The proof of the non-percolation}

It is essential for the proof that the potential function has the
hard core. However, the arguments in this subsection do not depend
on the dimension of the space.

The next lemma is an obvious consequence of the hard core
condition and the inequality \reff{tail}.
\begin{lemma}
Let $V\subseteq \R^\nu$ be a finite volume and let $\sigma$ be a
configuration in $V$. Then
\begin{equation}\label{exter}
\int_{V^c}e^{-H(\sigma\vee\omega)}\Pi(\ed \omega)<\infty.
\end{equation}
\end{lemma}

We obtain the non-percolation result by a coupling of the Gibbs
measure $P^{\beta,\lambda}$ with a {\it branching cluster
process}. The method was described in the work \cite{FPSY}. Here
we use the analogous idea.

Note that due to the hard core condition, one considers the
$\ell$-percolation  only when $\ell >f.$ Informally the idea of
the proof is the following. Suppose that $R \subset \omega$ is a
$\ell$-cluster with $\#(R)>1$. Let us choose some particle $x_0$
from this cluster, $x_0\in R.$ Let $R^{(1)} \subset R$ be the set
of all particles from $R$ such that the distance between $x_0$ and
any point from $R^{(1)}$ is less than or equal to $\ell.$ The set
$R^{(1)}$ is not empty because $\#(R)>1$. Next we choose a set
$R^{(2)},$ where $ R^{(2)}\subset R \setminus (R^{(1)} \cup
\{x_0\}),$ and which is the set of the particles such that for any
$v$ from $R^{(2)}$ there exists at least one point $w$ from
$R^{(1)}$ at the distance no greater than $\ell.$ We can call $v$
an {\it offspring} of $w$. If the set $R\setminus (R^{(2)}\cup
R^{(1)} \cup \{x_0\})$ is not empty we can choose a subset
$R^{(3)}$ with the similar properties, ect. Iterating the
procedure we will obtain the following representation of the
cluster $R:\  R = \cup_{i=0} ^\infty R^{(i)}$ (here
$R^{(0)}=\{x_0\}$). The set $R^{(i)}$ we call {\it $i$-th
generation}. The set $R^{(n-1)}$ generates the set $R^{(n)}.$

This branching construction brings us the idea of a branching
process, but there are two peculiarities that differ our process
from the ordinary branching process. First, note that it is
possible for one offspring to have different parents. Thus we do
not have here a branching tree and it means that the independence
of the offsprings does not hold. Second, keeping in mind the
coupling, we define the transition probabilities of the branching
process by Gibbs measure $P_V$. Thus for $\ell$ sufficiently small
it is possible that some generation $R^{(n)}$ interacts with
precedent generations $R^{(k)}, k<n.$

Next we describe a rigorous construction of the cluster branching
process.

\subsubsection{The cluster branching process}\label{coupl}
We describe a  path of the process and its distribution $\sf P$.

Let $x_0$ be some point from $\R^\nu$. We shall construct a
sequence $( R^{(n)})$ which describes the generations. Together
with the sequence $(R^{(n)})$ we define the sequence $(E_{n})$,
where $E_{n}=\cup_{i=0}^{n} R^{(i)}$ and the sequence of the
\textit{occupied areas} $B_n=\cup_{v\in E_{n-1}} B_\ell (v).$ The
set $E_n$ we call an \textit{environment}.

{\it Initial step.} Let $R^{(0)}=\{x_0\}$ and $E_{0} = \{x_0\}, \
B_0=\emptyset$.

{\it First step.} Let $R^{(1)}=\{x^{(1)}_1, \dots, x_{k_1}^{(1)}
\}$ be a set of the particles in the ball $B_\ell(x_0)$ with the
center $x_0$ and with the radius $\ell.$ The set $R^{(1)}$ is the
offspring set of $x_0.$ Then $E_1 = E_0 \cup R^{(1)}$ and
$B_1=B_\ell(x_0)$. $B_1$ is the occupied area by the offsprings of
$x_0$. No particles of further embranchments appear in $B_1$. We
define a conditional probability density $\rho$ of the measure
$\sf P$ with respect to the same Poisson measure $\Pi$. The
density with respect to $\Pi$ of the offspring set $R^{(1)}$ of
the ancestor $x_0$ is
\begin{equation} \label{off1}
\rho(x_1^{(1)}, \dots, x_{k_1}^{(1)} \mid E_0) = \frac{1}{Z(E_0)}
\int_{\Omega_{B_1^c}} e^{-\beta H(E_1 \vee \omega )}
\Pi(\ed\omega),
\end{equation}
where $\Omega_{B_1^c}$ is the set of all configurations where
particles ``live'' outside of the ball $B_1$, and
$$ Z(E_0) =
\int_{\Omega} e^{-\beta H(E_0 \vee \omega )} \Pi(\ed\omega).
$$
We use Gibbs measure $P^{\beta,\lambda}$ for the definition of
$\rho$. In fact, all defined probabilities and further
calculations assume a big volume $V$, where we consider all
configurations. It means that we use the measure $P_V$ instead
$P^{\beta,\lambda}$. For example, in \reff{off1} the integration
is taken over $\Omega_{B_1^c\cap V}$. Therefore $\rho$ depends on
$V$. However, in what follows all estimates do not depend on $V$,
and hence can be considered as the estimates in the infinite
volume. We do not mention the volume $V$ in the further
calculations except cases when it is required.

Using $\rho$ we can calculate, for example, the probability to
have $k$ offsprings of $x_0:$
\begin{eqnarray*}
{\sf P}(\#(R^{(1)}) =k \mid E_0) &=& \int_{\{ R^{(1)}\in
\Omega_{B_1} :\ \#(R^{(1)})=k \} } \rho( R^{(1)} \mid E_0) \Pi(\ed
R^{(1)}) \\ &=&  \frac{\lambda^{k}|B_1|^k}{k!} e^{- \lambda |B_1|}
\int_{(B_1)^k} \rho(x_1^{(1)}, \dots, x_{k}^{(1)} \mid E_0) \ed
x_1^{(1)}\dots \ed x_{k}^{(1)}.
\end{eqnarray*}

{\it Second step.} In this step we describe the embranchments of
all particles from $R^{(1)}$. Any particle branches according to
some order introduced in $R^{(1)}$. We shall construct the set
$R^{(2)}$ in according to the chosen order. Let $R^{(2,1)} =
\{x^{(2,1)}_{1}, \dots , x^{(2,1)}_{k_{2,1}} \} \subset B_\ell
(x_1^{(1)}) \setminus B_1$ be the set of  offsprings of
$x_1^{(1)}.$ The offsprings of $x_1^{(1)}$ cannot be situated in
$B_1$ since $B_1$ is occupied by the offsprings of $x_0$. Let
$E_{(2,1)} = E_1 \cup R^{(2,1)}$ and $B_{(2,1)} = B_1\cup B_\ell
(x_1^{(1)}).$ The probability density of the offsprings of
$x_1^{(1)}$ is
\begin{equation} \label{off2-1}
\rho(x_1^{(2,1)}, \dots, x_{k_{2,1}} ^{(2,1)} \mid E_1) =
\frac{1}{Z(E_1)} \int_{\Omega_{(B_{(2,1)})^c } } e^{-\beta
H(R^{(2,1)} \vee E_1 \vee \omega )} \Pi(\ed\omega),
\end{equation}
where
\begin{equation}\label{stat-1} Z(E_1)= \int_{
\Omega_{B_1^c}} e^{-\beta H(E_1 \vee \omega )} \Pi(\ed\omega)
\end{equation}
(see \reff{exter}).

Suppose now that $k<k_1$ particles from $R^{(1)}$ are already
branched and $R^{(2,1)}, \dots, R^{(2,k)}$ is a sequence of their
offsprings. Hence we have the environment (all already living
particles) $E_{(2,k)} = E_1 \cup \bigcup_{i=1}^{k} R^{(2,i)}$ and
the occupied area $B_{(2,k)} = B_1 \cup \bigcup_{i=1}^{k}
B_\ell(x_i^{(1)})$. Note that $B_{(2,k)}$ is the
$\ell$-neighborhood of $E_{0} \cup \{ x_1^{(1)}, \dots,
x_k^{(1)}\}$.

Let now the next point $x_{k+1}^{(1)}$ be branching. Let
$$
R^{(2,k+1)} = \left\{x^{(2,k+1)}_1, \dots ,
x^{(2,k+1)}_{k_{2,k+1}} \right\} \subset B_\ell (x_{k+1}^{(1)})
\setminus B_{(2,k)}
$$
be the set of  offsprings of $x_{k+1}^{(1)}$. Then $E_{(2,k+1)} =
E_{(2,k)} \cup R^{(2,k+1)}$ and $B_{(2,k+1)} = B_{(2,k)} \cup
B_\ell (x_{k+1}^{(1)})$. The probability density of the offsprings
is
\begin{equation} \label{off2-11}
\rho\left(x_1^{(2,k+1)}, \dots, x_{k_{2,k+1}} ^{(2,k+1)} \mid
E_{(2,k)}\right) = \frac{1}{Z(E_{(2,k)})} \int_{\Omega_{
(B_{(2,k+1)})^c} } e^{-\beta H(E_{(2,k+1)} \vee \omega )}
\Pi(\ed\omega).
\end{equation}

We obtain the next generation $R^{(2)} = \bigcup_{i=1}^{k_1}
R^{(2,i)}$ after the embranchments of all points from $R^{(1)}$.
Let $E_2 = E_{(2,k_1)}$ and $B_2 = B_{(2,k_1)}$.

{\it $(n+1)$-th step.} To construct $R^{(n+1)}$ from $R^{(n)}$ we
 follow the same scheme. Let $R^{(n)} = \{ x_1^{(n)}, \dots,
x_{k_n}^{(n)} \}$. The particles from $R^{(n)}$ are branching
according to some chosen order in $R^{(n)}$. Suppose that  $k$
($k< k_n$) particles from $R^{(n)}$ are branched. Hence we have
the sets $R^{(n+1,i)}$ (where $i\le k$), $E_{(n+1,k)}$ and
$B_{(n+1,k)}$.

Now, let
$$
R^{(n+1,k+1)} = \left\{ x_1^{(n+1,k+1)},\dots, x_{k_{n+1,k+1}}
^{(n+1,k+1)} \right\} \subset B_\ell(x_{k+1}^{(n)}) \setminus
B_{(n+1,k)}
$$
be the set of  offsprings of the branching particle
$x_{k+1}^{(n)}$. Then $E_{(n+1,k+1)} = E_{(n+1,k)} \cup
R^{(n+1,k+1)}$ and $B_{(n+1,k+1)} = B_{(n+1,k)} \cup
B_\ell(x_{k+1} ^{(n)})$. The probabilistic density is
\begin{eqnarray} \nonumber
&& \rho\left(x_1^{(n+1,k+1)},\dots,
x_{k_{n+1,k+1}} ^{(n+1,k+1)} \mid E_{(n+1,k)} \right) \\
\label{off2-n} && \phantom{x_1^{(n+1,k+1)}} {} =
\frac{1}{Z(E_{(n+1,k)})} \int_{\Omega_{ (B_{(n+1,k+1)})^c} }
e^{-\beta H(E_{(n+1,k+1)} \vee \omega )} \Pi(\ed\omega).
\end{eqnarray}

The above iterative steps describe a path and the transition
probabilities of the cluster growth process. Note that the
offsprings can depend not only on the preceding  generation, but
on {\it all} previous generations.

The question we are interested is when the cluster branching
process extincts with the probability 1. It is known that the
extinction condition of the ordinary branching processes is
formulated in the term of the mean number of the offsprings. We
can expect that the extinction of the cluster branching process is
controlled by the offspring mean value, as well. We show further
that if the mean number of the offsprings is uniformly less than 1
over all possible cluster configurations of the previous
generations, then the cluster branching process will extinct with
the probability 1.

In the next lemma we prove that there exists a region in the plane
$(\lambda, \beta)$ such that the mean offspring number of one
ancestor is less than 1. Then, we show that this condition is
sufficient for the extinction of the cluster branching process.
Moreover we show that the mean value of  paths of the cluster
branching process is finite.

\begin{lemma} \label{less1}
There exists $\lambda_\ell^-$ and a function $\beta^-(\lambda)$
such that for any $\lambda<\lambda_\ell^-$ and $\beta <
\beta^-(\lambda)$ the expected number of the offsprings
$\#(R^{(n,k)})$ of the ancestor $x_k^{(n-1)}$  is less than 1,
uniformly over $n, k$ and over the environment $E_{(n,k-1)}.$
\begin{equation} \label{mean-off}
\sup_{n,k}{\sf E} \bigl( \#( R^{(n,k)} ) \ \bigl|\bigr.
E_{(n,k-1)} \bigr) < 1.
\end{equation}
\end{lemma}

\pr We give an estimate of the probability the point
$x_{k}^{(n-1)}$ to have exactly $K$ offsprings, that is
$\#(R^{n,k})=K$. By the definition of the offspring density
\reff{off2-n} we have to estimate the following integral. Let $
\wt B ( x_{k}^{(n-1)}) := B_\ell ( x_{k}^{(n-1)}) \setminus
B_{(n,k-1)} $ then
\begin{eqnarray}
&& {\sf P} \bigl( \#( R^{(n,k)} ) = K \ \bigl|\bigr.\ E_{(n,k-1)}
\bigr) \label{probk}\\
&& {} = \int_{ \left\{ R^{(n,k)} \in\Omega_{ \wt B \left(
x_{k}^{(n-1)} \right) }: \#(R^{(n,k)}) = K  \right\} }
\rho(R^{(n,k)} \mid E_{(n,k-1)}) \Pi(\ed R^{(n,k)} ) \nonumber
\end{eqnarray}
We shorten some notations in the further calculations. Let  $\wt R
:= R^{(n,k)}$ and $\wt E := E_{n}^{(k-1)}$. Than the integral
\reff{probk} can be represented as
\begin{eqnarray} \label{H1}
&& {\sf P} ( \#( R^{(n,k)}) = K \mid E_{(n,k-1)} ) = {\sf P} ( \#(
\wt R) = K \mid \wt E )  \\ && {} = \frac{1}{Z(\wt E)} \int_{
\bigl\{ \wt R \in\Omega_{ \wt B \left( x_{k}^{(n-1)} \right) }:
\#(\wt R) = K  \bigr\} } \int_{\Omega_{B_{(n,k)}^c} } e^{-\beta H(
\wt R \vee \wt E \vee \omega )} \Pi(\ed\omega) \Pi(\ed\wt R),
\nonumber
\end{eqnarray}
where
\begin{equation} \label{stat1}
Z(\wt E) = \int_{\Omega_{B_{(n,k)}^c}} e^{ - \beta H(\wt E \vee
\omega) } \Pi(\ed\omega).
\end{equation}
Hamiltonian in \reff{H1} can be represented  as
\begin{equation}\label{H2}
H(\wt R \vee \wt E \vee \omega) = H(\wt R) + H(\wt E \vee \omega )
+ F(\wt R, \wt E \vee \omega).
\end{equation}
Let ${\sf C}_m$ be a sequence of the strips
$$
{{\sf C}_m} =\{ x \in \mathbb R^\nu:\ m\le |x - x_k^{(n-1)}|< m+1
\},
$$
where $m\in \Z_+$ and $m\geq g$, and
$$
{{\sf C}_0}=\{ x \in \mathbb R^\nu:\ \ell\le |x - x_k^{(n-1)}|<
m_0 \},
$$
where $m_0=\min\{m:\:m\geq g\}$. The volume of ${\sf C}_m$ is
$$
|{\sf C}_m|=\kappa P_{\nu-1}(m),
$$
where $P_{\nu-1}(r)$ is a polynomial having its power $\nu-1$, and
the leading coefficient is $\nu-1$. Let $\Omega_m=\Omega_{{\sf
C}_m\setminus B_{(n,k)}}.$ Let $\omega\in \Omega_m$ then the
number of the particles of the configuration $\omega\vee \wt{E}$
in ${\sf C}_m$ has the following bound
$$
\#\left((\omega\vee \wt{E})_{{\sf C}_m}\right)\leq
\left[\frac{|{\sf
C}_m|}{\kappa\left(\frac{f}{2}\right)^\nu}\right]\leq
\frac{P_{\nu-1}(m)}{\left(\frac{f}{2}\right)^\nu},
$$
where $(\omega\vee \wt{E})_{{\sf C}_m}$ is the restriction of
$\omega\vee \wt{E}$ to ${\sf C}_m$. We have then the following
lower bound for the interaction energy of $(\omega\vee
\wt{E})_{{\sf C}_m}$ and $\wt{R}$ consisting $K$ offsprings,
$\#(\wt{R})=K$,
$$
F\left((\omega\vee \wt{E})_{{\sf C}_m},\wt{R}\right)\geq
-\psi(m)K\frac{2^\nu P_{\nu-1}(m)}{f^\nu}.
$$
The energy of the interaction of $\omega\vee \wt{E}$ and $\wt{R}$
is estimated as
\begin{align*}
F\left(\omega\vee \wt{E},\wt{R}\right)&\geq&& -KMn_0-
\frac{2^\nu K}{f^\nu}\sum_{m=m_0}^\infty P_{\nu-1}(m)\psi(m)\\
&\geq&&-KMn_0-\frac{2^\nu K}{f^\nu}\int_{m_0}^\infty P_{\nu-1}(r)\psi(r)\ed r\\
&\geq&&-K\left(Mn_0+\frac{2^\nu}{f^\nu}I\right),
\end{align*}
where $n_0=\left[\frac{|{\sf
C}_0|}{\kappa\left(\frac{f}{2}\right)^\nu}\right]$ (see
\reff{tail}). Let $n_1=Mn_0+\frac{2^\nu}{f^\nu}I$.

By \reff{H1} and \reff{H2} we have
\begin{eqnarray*}
&&
{\sf P} ( \wt R: \#(\wt R) = K \mid \wt E ) \\
&& {} \le \frac{e^{\beta K n_1 }}{Z(\wt E)} \int_{ \bigl\{ \wt R
\in\Omega_{ \wt B \left( x_{k}^{(n-1)} \right) }: \#(\wt R) = K
\bigr\} } \!\!\!\!\!\!\!\! e^{ -\beta H(\wt R)} \Pi(\ed \wt R)
\int_{\Omega_{B_{(n,k)}^c} } e^{ - \beta H(\wt E \vee \omega) }
\Pi(\ed \omega).
\end{eqnarray*}
The cluster energy $H(\wt R)$ we can estimate very roughly as
\begin{equation} \label{2}
-H(\wt R) < M \binom{K}{2} < MK^2.
\end{equation}
Thus
\begin{eqnarray}
&&
{\sf P} ( \wt R: \#(\wt R)=K \mid \wt E ) \nonumber \\
&& {} \le \frac{e^{\beta (K n_1 + M K^2)}}{Z(\wt E)} \int_{
\bigl\{ \wt R\in\Omega_{ \wt B \left( x_{k}^{(n-1)} \right) }:
\#(\wt R) = K \bigr\} } \Pi(\ed \wt R) \int_{\Omega_{B_{(n,k)}^c}
}  e^{ - \beta H(\wt E \vee \omega) } \Pi(\ed \omega) \nonumber
\\
&& {} \le \frac{e^{\beta (Kn_1 + M K^2)}}{Z(\wt E)} \frac{(\lambda
| \wt B ( x_{k}^{(n-1)})  |)^K }{K!} e^{ - \lambda | \wt B (
x_{k}^{(n-1)})  | } \int_{\Omega_{B_{(n,k)}^c} }
 e^{
- \beta H(\wt E \vee \omega) } \Pi(\ed\omega) \nonumber
\\
\label{3} && {} \le e^{\beta (Kn_1 + M K^2)} \frac{(\lambda
|B_\ell (x_{k}^{(n-1)}) |)^K }{K!}.
\end{eqnarray}
In the last inequality we use
$$
\frac{1}{Z(\wt E)} \int_{\Omega_{ B_{(n,k)}^c} } e^{ - \beta H(\wt
E \vee \omega) } \Pi(\ed \omega) = 1.
$$
Besides in the last inequality in \reff{3} we estimate the volume
$|\wt B( x_{k}^{(n-1)})|$ by $|B_\ell ( x_{k}^{(n-1)})|$. Note
that $\wt B ( x_{k}^{(n-1)})$ can be empty, that means that the
particle $x_{k}^{(n-1)}$ has no offsprings. It leads to zero of
the probability we are estimating. We do not use this possibility.

The right hand side of \reff{3} does not depend on the volume $V$.

The maximal number of the particles in the ball $B_\ell
(x_{k}^{(n-1)})$ is $n_B=|B_\ell (x_{k}^{(n-1)}) |/(\kappa
(f/2)^\nu).$ Thus using estimation \reff{3} we can estimate the
mean number of the offsprings:
\begin{eqnarray} \nonumber
{\sf E} \bigl( \#(\wt R) \ \bigl|\bigr. \  \wt E \bigr) &\le&
\sum_{k=0}^{n_B} k e^{\beta kn_1 + \beta k^2 M} \frac{(\lambda
|B_\ell (x_{k}^{(n-1)}) |) ^{k} } {k!}  \\
\label{mean}& \le &  \lambda \kappa \ell^\nu  e^{\beta n_Bn_1 +
\beta n_B^2 M}  e^{\lambda \kappa \ell^\nu}.
\end{eqnarray}
Let
\begin{equation}\label{b-minus}
\beta^-_\ell(\lambda) =  - \frac{1}{A} \ln \lambda -\frac{\kappa
\ell^\nu}{A} \lambda - \ln\left(\frac{\kappa \ell ^\nu
}{A}\right),
\end{equation} where
$$
A= n_B  (Mn_B+n_1).
$$
The function $\beta^-(\lambda)$ is considered on the interval $(0,
\lambda_\ell^-]$, where $\lambda_\ell^-$ is a root of the equation
\begin{equation} \label{12}
-  \ln \lambda -{\kappa \ell^\nu} \lambda - \ln({\kappa \ell
^\nu}) =0.
\end{equation}
When $\lambda<\lambda_\ell^-$ and $\beta < \beta^-(\lambda)$ we
obtain
\begin{equation}\label{less}
{\sf E} \bigl( \#( \wt R) \ \bigl|\bigr.\  \wt E \bigr) < 1
\end{equation}
uniformly over the environment $\wt E.$ $\Box$

Next we show that if the mean number of the offsprings is less
than 1 (see \reff{less}) then the mean size of the cluster is
finite. Indeed, if $\beta < \beta^-(\lambda)$, then there exists
$\epsilon>0$ depending on $\beta$ and $\lambda$ such that ${\sf
E}( \#(\wt R) \mid \wt E) < 1 -\epsilon$, and
\begin{eqnarray*}
{\sf E} ( \#(R^{(n)}) ) &= & {\sf E} \Bigl( \sum_{k=1} ^\infty I_{
\{ \#(R^{(n-1)})=k\} }
\#(R^{(n)}) \Bigr) \\
&=& \sum_{k=1} ^\infty \sum_{i=1} ^k  {\sf
E} \Bigl( I_{ \{ \#(R^{(n-1)})=k\} } \#(R^{(n,i)}) \Bigr) \\
&=& \sum_{k=1} ^\infty \sum_{i=1} ^k  {\sf E} \Bigl( I_{
\{\#(R^{(n-1)})=k\} } {\sf
E} \bigl( \#(R^{(n,i)}) \bigl|\bigr. E_{(n,i-1)} \bigr) \Bigr) \\
&\le & (1-\epsilon) \sum_{k=1} ^\infty \sum_{i=1} ^k  {\sf E} (
I_{ \{ \#(R^{(n-1)}) = k \} } ) = (1-\epsilon) {\sf E} (
\#(R^{(n-1)}) ).
\end{eqnarray*}
It means that
$$ {\sf E} ( \#(R^{(n)}) ) < (1-\epsilon)^n, $$
and we see that the mean cluster size is finite,
\begin{equation}\label{12.1}
{\sf E} \Bigl( \sum_{n=1}^\infty \#(R^{(n)}) \Bigr) = \sum_{n=1}
^\infty E( \#( R^{(n)}) ) \le 1/\epsilon.
\end{equation}
Since the path $R^{(n)}$ is a $\ell$-connected set of points in
$V$ and the estimate \reff{12.1} does not depend on $V$, the
probability of infinite clusters is 0.

This proves that the cluster branching process is degenerated.

Any path of the cluster branching process starting from $x_0$ is
$$
E =  \cup_{n=1} ^ N R^{(n)},
$$
where $N$ is the number of the generations. $N$ is finite since
\reff{12.1}. The relation \reff{12.1} can be rewritten as
\begin{equation}\label{12.2}
{\sf E} (\#(E)) = \sum_{k} k \int_{ \{ E:\ \#(E)=k \} } \rho(E
\mid E_0) \Pi(\ed E) < \infty.
\end{equation}

\subsubsection*{The coupling} We explain next the coupling of the Gibbs
field having the distribution $P^{\beta,\lambda}$ and the
branching cluster process having the distribution $\sf P$. To this
end we represent any finite $\ell$-cluster of a configuration
$\omega$ as a path of the branching cluster process.

Let $\gamma \subset \omega_0 \in \Omega$ be a finite
$\ell$-cluster in a configuration $\omega_0$. It means that
$B_{\ell/2} (\gamma) = \cup_{x\in \gamma} B_{\ell/2} (x)$ is
connected component in $B_{\ell/2}(\omega_0) = \cup_{x\in
\omega_0} B_{\ell/2} (x).$

Choose a particle $x_0 \in \gamma$. We can consider the
configuration $\gamma$ as a branching process path $E=(E_n),\ E_n
\subseteq E_{n+1}$, starting at the particle $x_0\; (E_0 = \{ x_0
\})$, and such that $\gamma=\cup E_n$. The construction was
described in the section \ref{coupl}. By the described iteration
we obtain also a sequence of generations $(R^{(n)})$ such that
$\gamma = \cup_{n} R^{(n)}$. Any generation $R^{(n)}$ is a set of
the offsprings of $R^{(n-1)}$.

Next we find a relation between the distribution of the cluster
branching process and the Gibbs measure. Consider the event
$$
\Delta_\gamma = \{ \omega: \ \gamma \subset \omega,\ B_\ell
(\gamma) \cap \gamma = \gamma \}.
$$
We can define a density $\chi$ of this event with the respect to
$\Pi$.

 \begin{equation} \label{chi}
 \chi(\Delta_\gamma) = \frac{1}{ Z_V} \int_{ \Omega_{ ( B_\ell (\gamma) )^{c} }}
 e^{ -\beta H(\gamma \vee \omega)} \Pi(\ed\omega)
 \end{equation}
 The density $\rho(\gamma)$ of the path $\gamma$ is
\begin{equation}\label{rho}
\rho(\gamma \mid E_0) = \prod_{n=1} ^N \rho( R^{(n)} \mid
E_{n-1}).
\end{equation}
It follows from finiteness of $\gamma$ that $N$ is finite.

It is not difficult to verify that
 \begin{equation}\label{rho-1}
\rho(\gamma \mid E_0) = \frac{ Z( E_N) }{ Z(E_0)}. \end{equation}
Then
\begin{equation} \label{chi-rho}
\chi(\Delta_\gamma) = \rho(\gamma \mid E_0) \frac{ Z(E_0)}{Z_V}.
\end{equation}

Let $(V_m)$ and $(\wt V_m)$ be sequences of the boxes
$$
V_m = \{ x\in \mathbb R^\nu:\ |x|\le m \},\ \ \wt V_m = \{ x\in
\mathbb R^\nu:\ |x|\le m-\ell \}.
$$
Define the sequence $(\gamma_m)$ where $\gamma_m = \gamma \cap \wt
V_m$. For any $\gamma_m$ we have the relation  \reff{chi-rho}.

Finiteness of the expectation of the $\ell$-clusters follows from
\reff{chi-rho}. Let
$$
\Gamma_{x_0} ^k = \{ \gamma \subset \mathbb R^\nu:\ \#(\gamma)=k,
\ x_0\in \gamma\}
$$
be the set of all $\ell$-clusters containing the particle $x_0$
and having exactly $k$ particles.

Consider the set of the configurations
$$
\Omega^{\Gamma_{x_0} ^k} = \bigcup_{\gamma \in \Gamma_{x_0} ^k}
\Delta_\gamma
$$
and its Gibbs probability
\begin{eqnarray*}
P_V \bigl( \Omega^{\Gamma_{x_0} ^k} \bigr) &=& \frac{1}{Z_V}
\int_{ \Omega^{\Gamma_{x_0} ^k} } e^{-\beta H(\omega) } \Pi(\ed \omega) \\
&=&   \frac{1}{Z_V} \int_{ \Gamma_{x_0} ^k } \int_{ \Omega_{\left(
B_\ell(\gamma)\right)^c} } e^{-\beta H(\omega \vee \gamma) }
\Pi(\ed \omega) \Pi(\ed ( \gamma\vee \phi_{ B_\ell(\gamma)} ) ),
\end{eqnarray*}
where $\phi_{ B_\ell(\gamma)}$ is the empty configuration in the
$\ell$-neighborhood of $\gamma$. Hence
\begin{equation*}
P_V \bigl( \Omega^{\Gamma_{x_0} ^k} \bigr) = \int_{ \Gamma_{x_0}
^k } \rho( \gamma \mid x_0) \Pi( \ed( \gamma \vee \phi_{
B_\ell(\gamma)} ) )
\end{equation*}

   The mean value of the size of the clusters $\gamma$ then is
\begin{eqnarray*}
E_V (\#(\gamma)) &=& \sum_{k=1}^\infty k P_V \bigl(
\Omega^{\Gamma_{x_0} ^k} \bigr) \\
&=& \sum_{k=1}^\infty k \int_{ \Gamma_{x_0} ^k } \rho( \gamma \mid
x_0) \Pi( \ed( \gamma \vee \phi_{ B_\ell(\gamma)} ) ) < \infty
\end{eqnarray*}
   since \reff{12.2}. This proves Theorem~\ref{hard-core} \qed

\subsection{The proof of the percolation }\label{proof}The proof of
the existence of infinite clusters is based on technics which is
close to the contour method in the lattice  models. To apply the
method we discretize $\R^2$ splitting it into squares. A
$c$-contour around 0 is a set of empty (without particles) squares
surrounding 0. The main fact we prove is that the probability of a
$c$-contour decreases exponentially with its length. It leads to
the finiteness of the number of the contours surrounding 0.

In our proof of the percolation, the essential  assumption is that
the space is two-dimension. The hard core is not used.

Divide $\mathbb R^2$ into square cells
$\mathcal{S}=\{S_{(k,l)}^{q}\}$ of the linear size equal to $q.$
Suppose that the left-lower corner  of any cell $S_{(k,l)}^{q}$
has coordinate $(kq, lq),$ where $(k,l) \in \mathbb Z^2.$ So we
have a natural order of the cells.  The point
$c_{(k,l)}=\left(\frac{2k+1}{2}q,\frac{2l+1}{2}q\right)$ is called
the center of the cell  $S_{(k,l)}^{q}$. Two cells $S_{(k,l)}^{q}$
and $S_{(k',l')}^{q}$ are neighbors if either $k=k'\pm 1$ and
$l=l'$ or $l=l'\pm 1$ and $k=k'$. Let $\langle c,c'\rangle$ be the
line connecting the centers $c=c_{(k,l)}$ and $c'=c_{(k',l')}$ if
$S_{(k,l)}^{r}$ and $S_{(k',l')}^{q}$ are neighbors. Let
$\mathcal{P}=\{S_{(k,l)}^{q}\}\subseteq \mathcal{S}$ be a finite
subset of the cells and
$\mathbf{C}(\mathcal{P})=\{c_{(k,l)}:\:S_{(k',l')}^{q}\in
\mathcal{P}\}$ be the set of all centers of the cells from
$\mathcal{P}$. For every set $\mathcal{P}$ of the cells we
consider the graph
$$
G_{\mathcal{P}}=\Big(\mathbf{C}(\mathcal{P}),
\Gamma(\mathcal{P})=\{\langle c,c'\rangle:c,c'\in
\mathbf{C}(\mathcal{P}) \} \Big)
$$
having
$\mathbf{C}(\mathcal{P})$ as the vertex set and
$\Gamma(\mathcal{P})$ as the bond set of all bonds connecting
neighboring cells from $\mathcal{P}$. A set of cells $\mathcal{P}$
is connected if the graph $G_{\mathcal{P}}$ is connected.

A set of cells $\mathcal{R}$ is called \textit{contour} if the
bond set $\Gamma(\mathcal{R})$ is homeomorphic to the circle. The
number $n(\mathcal{R})$ of the cells in a contour $\mathcal{R}$ is
called the length of the contour.

If $\mathcal{P}$ is a set of cells then
$W(\mathcal{P})=\bigcup_{S\in\mathcal{P}}S\subseteq\R^{2}$ is the
support of $\mathcal{P}$.

All contours we consider further surround $0\in\R^2$. Therefore we
often omit mentioning this. Let $\omega\in\Omega$ be a
configuration. If a contour $\mathcal{R}$ is such that $\omega\cap
W(\mathcal{R})=\varnothing$ then we call it a $c$-\textit{contour
with the respect to }$\omega$ or simply a $c$-\textit{contour}.

The proof of Theorem \ref{t2} is based on the following lemma. Let
$\Omega^0(\mathcal{R})$ be the event (the set of configurations)
such that the contour $\mathcal{R}$ is the $c$-contour with
respect to any $\omega \in \Omega^0(\mathcal R)$, that is
$\Omega^{0}(\mathcal{R})=(\omega_{W( \mathcal R)}=\varnothing).$
Let $\wt{\Omega}=\Omega_{W(\mathcal R)^c}$ be the set of all
configurations out of $W(\mathcal R)$. Then
$$
\Omega^{0}(\mathcal{R})=\{\omega=\phi_{W(\mathcal
R)}\vee\wt{\omega}:\:\wt{\omega}\in\wt{\Omega}\},
$$
where $\phi_{W(\mathcal R)}$ is the empty configuration in
$W(\mathcal R)$.

\begin{lemma}\label{l1}
Let the cell size be $q=2d+\delta,$ where $\delta$ is a small
positive number.  There exist constants $\alpha\in(0,1)$,
$c(\beta)>0$ and $G(\beta,\lambda)$ such that for any  $h\geq 0$
there exists a function ${\beta}_{\ell,h}^+(\lambda)\geq 0$
defining the domain
$$
\{(\lambda,\beta):\:\beta>{\beta}_{\ell,h}^+(\lambda)\}
$$
where the following relations hold:
\begin{enumerate}
    \item $G(\beta,\lambda)>h$,
    \item for any $c$-contour $\mathcal{R}$
\begin{equation} \label{mainlemma}
P^{\beta,\lambda}(\Omega^{0}(\mathcal{R})) < c(\beta)
e^{-n(\mathcal{R}) \alpha G\beta,\lambda)}.
\end{equation}
The probability that there are no particles in set $W(\mathcal R)$
which is the support of the $c$-contour $\mathcal{R}$
exponentially decreases with the contour length.
\end{enumerate}
\end{lemma}

\pr Let $V$ be a volume in $\R^{2}$ containing
$W(\mathcal{R})=\bigcup_{S\in\mathcal{R}}S$.  In order to estimate
the probability of event $\Omega^0(\mathcal{R})$ (see
Figure~\ref{fig2}-A) we  construct a event $\Omega^1(\mathcal{R})$
by adding particles in the $c$-contour $\mathcal R$     (see
Figure~\ref{fig2}-B).  That allows to obtain the lower bound of
probability $P_V(\Omega^1(\mathcal{R}))$ of the form
$P_V(\Omega^1(\mathcal{R}))>e^{ n\alpha G(\beta,\lambda)} P_V(
\Omega ^0)$, where $n=n(\mathcal{R})$.   Substituting the
probability $P_V(\Omega^1(\mathcal{R}))$ by 1, we immediately
obtain (\ref{mainlemma}).

\subsubsection*{\bf The probability of $\Omega^0(\mathcal{R})$}
Recall that $\Omega^{0}(\mathcal{R})$ is the event composed of the
configurations in $V$ containing $c$-contour $\mathcal{R}$. We
assume that the boundary configuration out of $V$ is
$\tau=\varnothing$. The probability of the event then is
\begin{eqnarray}\label{omega0}
P_{V}(\Omega^{0}(\mathcal{R}))
&=&\frac{1}{Z_{V}}\int_{\Omega^{0}(\mathcal {R} )} \exp\{-\beta
H(\omega)\}\Pi(\ed \omega) \\ &=& \frac{e^{-\lambda\Delta n
}}{Z_{V}}\int_{\wt{\Omega}}\exp\{-\beta H(\wt{\omega})\}\Pi(\ed
\wt{\omega}), \nonumber
\end{eqnarray}
where $\Delta=(2d+\delta)^2$ is the volume of any cell,
$\wt{\Omega}$ is the set of all configurations in $V\setminus
W(\mathcal{R})$.

Let $\phi_{W(\mathcal{R})}$ be the empty configuration in the
region $W(\mathcal{R})$. Any configuration
$\omega\in\Omega^0(\mathcal{R})$ is the composition of
$\phi_{W(\mathcal{R})}$ and a configuration $\wt{\omega}$ in $V
\setminus W(\mathcal{R})$,
$\omega=\phi_{W(\mathcal{R})}\vee\wt{\omega}$. The pre-integral
factor in \reff{omega0} is the integration result over
$\phi_{W(\mathcal{R})}$.

\subsubsection*{The construction of the event $\bf \Omega^1(\mathcal{R})$}
\begin{figure}
\begin{center}
\includegraphics[scale=0.5]{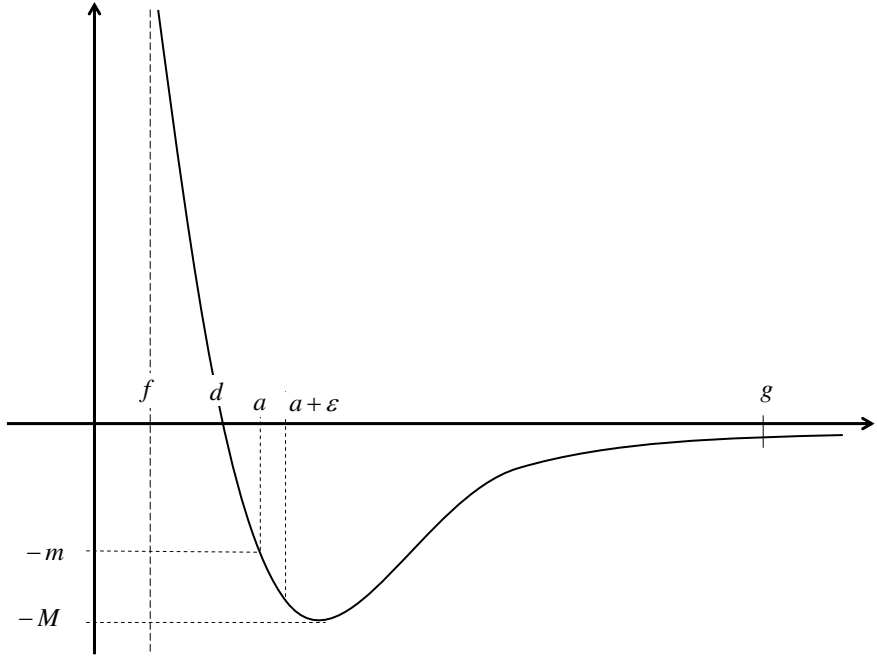}
\end{center}
\caption{The potential function.}%
\label{fig3}
\end{figure}

\begin{figure}
\begin{center}
\includegraphics[scale=0.5]{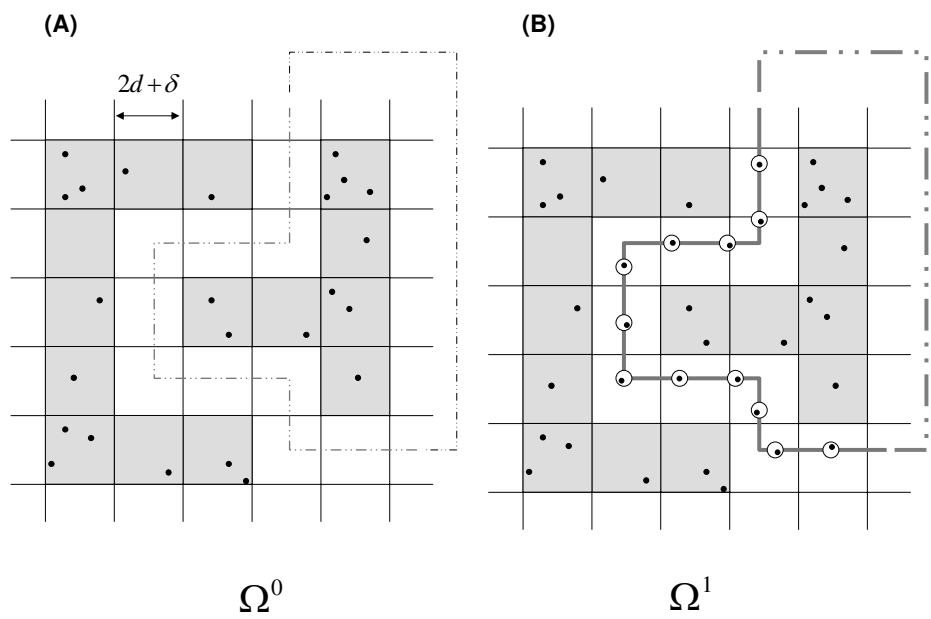}
\end{center}
\caption{The region of the positivity of the percolation function.}%
\label{fig2}
\end{figure}

Let $m$ be a positive number such that $m<M.$ For such $m$ there
exist positive numbers $a$ and $\varepsilon$ such that
$\varepsilon\leq \delta$ and
$$\varphi(x)\le -m\ \mbox{for all }|x|\in[a, a+\varepsilon],$$
(see Figure~\ref{fig3}.)

Let ${\gamma}=\bigcup_{\langle
c,c'\rangle\in\Gamma(\mathcal{R})}\langle c,c'\rangle$ be the line
in $\R^{2}$ composed of the bonds $\langle
c,c'\rangle\in\Gamma(\mathcal{R})$. The length of $\gamma$ is
equal to $(2d+ \delta) n(\mathcal{R}).$ We introduce a direction
on $\gamma$. Let it be counterclockwise. The points of any finite
subset $\wt{D}$ of $\gamma$ we numerate along the direction of
$\gamma$, that is $\wt{D}=\{x_0,...,x_k\}$. The point
$x_{i+1}\in\wt{D}$ is the first point which can be reached from
$x_i$ in the direction of $\gamma$.

Let $B_{a+\frac{\varepsilon}{2}}(x_i)$ be the closed disc of the
radius $a+\frac{\varepsilon}{2}$ with its center at
$x_i\in\wt{D}$. If $a+\frac{\varepsilon}{2}<4d+2\delta$ then
$B_{a+\frac{\varepsilon}{2}}(x_i)\cap\gamma$ is connected. In the
case $a+\frac{\varepsilon}{2}\geq 4d+2\delta$ the set
$B_{a+\frac{\varepsilon}{2}}(x_i)\cap\gamma$ can be unconnected.
Let $C(x_i)$ be a connected component of
$B_{a+\frac{\varepsilon}{2}}(x_i)\cap\gamma$ containing $x_i$.

Now we consider $\wt{D}$ only such that $x_{i+1}\in C(x_i)$ and
$|x_{i+1}-x_i|=a+\frac{\varepsilon}{2}$ for any $i<k$. Every  pair
$x,y\in\wt{D}$ such that $|x-y|=a+\frac{\varepsilon}{2}$ we call
\textit{connected}. The number $\eta(\wt{D})$ of the connected
pairs in $\wt{D}$ takes one of two values
$$
\#(\wt{D})-1\leq \eta(\wt{D})\leq \#(\wt{D}).
$$

The number $\#(\wt{D})$ of points in any $\wt{D}$ is not greater
than $\left[\frac{(2d+ \delta)
n(\mathcal{R})}{a+\frac{\varepsilon}{2}}\right]$.

Let $D=D(\mathcal{R})$ be such that
$\#(D)=\sup_{\wt{D}}\#(\wt{D})$.

An inverse estimate is in the next
\begin{lemma} There exists $\alpha > 0$
such that for any contour $\mathcal{R}$ the number
\begin{equation} \label{estB}
\#(D(\mathcal{R}))\geq \alpha n(\mathcal{R}) \end{equation} if
$n(\mathcal{R})>\frac{2\sqrt{2}a}{d}$.
\end{lemma}
\pr We consider two different cases distinguished by the following
inequalities
\begin{description}
    \item[case 1.] $a+\frac{\varepsilon}{2}<4d+2\delta$,
    \item[case 2.] $a+\frac{\varepsilon}{2}\geq 4d+2\delta$
\end{description}
The length $s(x_i,x_{i+1})$ of the piece of
$C(x_i)\subseteq\gamma$ between $x_i$ and $x_{i+1}$ can be
estimated as following
\begin{equation*}
s(x_i,x_{i+1})\leq\begin{cases}\sqrt{2}
\left(a+\frac{\varepsilon}{2}\right)&\mbox{ in the case 1.},\\
\left(a+\frac{\varepsilon}{2}\right)
\left(\frac{a+\frac{\varepsilon}{2}}{2d+\delta}+1\right) &\mbox{
in the case 2.}
\end{cases}
\end{equation*}
Since
$$
\#(D)\geq
\frac{n(\mathcal{R})\left(a+\frac{\varepsilon}{2}\right)}{s(x_i,x_{i+1})}
$$
we can take
\begin{equation*}
\alpha\geq\begin{cases} \frac{1}{\sqrt{2}} &\mbox{ in the case 1.}\\
\frac{2d+\delta}{a+\frac{\varepsilon}{2}+2d+\delta }&\mbox{ in the
case 2.}
\end{cases}
\end{equation*}
\qed

\vspace{.3cm}

Let $B_{\frac{\varepsilon}{4}}(x)$ be the disc of the radius $
\frac{\varepsilon}{4}$ centered at $x\in D$ and $U=\cup_{x \in D}
B_{\frac{\varepsilon}{4}}(x)$. Every disc
$B_{\frac{\varepsilon}{4}}(x)$ is called a {\textit{bead}} and the
set $U$ is {\textit{necklace}}. The set
\begin{equation*}
\Sigma_U=\{\sigma\in \Omega_{U}:\:\forall x\in D,\  \#(\sigma\cap
B_{\frac{\varepsilon}{4}}(x))=1, \sigma\cap U^c=\varnothing\}
\end{equation*}
is a set of configurations all particles of which are located in
the beads only, one particle in every bead.

The configuration set $\Omega^{1}(\mathcal{R})$ contains
configurations composed by the joint of three configurations:
\begin{equation}\label{om1}
\Omega^{1}(\mathcal{R})=\{\omega_1=\sigma\vee
\phi_{W(\mathcal{R})\setminus
U}\vee\wt{\omega}:\:\sigma\in\Sigma_U, \wt{\omega}\in\wt{\Omega},
\}
\end{equation}
where $\phi_{W(\mathcal{R})\setminus U}$ is the empty
configuration in $W(\mathcal{R})\setminus U$ and $\wt{\omega}\in
\wt{\Omega}$ are configurations in $W(\mathcal{R})^c$.

\subsubsection*{ The lower bound for $\bf P^{\beta,\lambda}
\left(\Omega^1(\mathcal{R})\right)$}

The probability of $\Omega^{1}(\mathcal{R})$ is
\begin{eqnarray*}
P_{V}( \Omega^{1}(\mathcal{R})) &=& \frac{1}{Z_{V}}
\int_{\Omega^{1} } e^{-\beta H_{V}(\sigma )} e^{-\beta
H_{V}(\wt{\omega})} e^{-\beta F(\sigma,\wt{\omega})}
\Pi(\ed(\sigma\vee\phi_{W(\mathcal{R})\setminus
U}\vee\wt{\omega})).
\end{eqnarray*}
The energy $F(\sigma,\tilde\omega)$ of the interaction of $\sigma$
and $\wt{\omega}$ is negative because the distance between any
particles of $\wt{\omega}$ and of $\sigma$ is greater than $d,$
hence $e^{-\beta F(\sigma, \wt{\omega})}$ is greater than 1. Since
$\phi_{W(\mathcal{R})\setminus U}=\varnothing,$ then
\begin{equation}\label{fe}
P_{V}(\Omega^{1}(\mathcal{R})) \geq  \frac{e^{-\lambda
n\Delta}e^{\lambda \frac{\pi \varepsilon^2}{16}\#(D)}}{Z_{V}}
\int_{\Sigma_{U}} e^{-\beta H_{V}(\sigma )}\Pi(\ed\sigma)
\int_{\wt{\Omega}}e^{-\beta H_{V}(\wt{\omega})}
\Pi(\ed\wt{\omega}).
\end{equation}

To estimate $H_V(\sigma)$ remark that there exist at least
$\#(D)-1$ the connected pairs in $D$. The interaction energy of
any connected pair $x,y\in D$ is estimated from below as
\begin{equation*}
\varphi(x-y)\leq -m.
\end{equation*}
Other pairs $x,y\in D$ which are not connected, that is $|x-y|\neq
a+\frac{\varepsilon}{2}$, interact with a negative energy.

Hence
\begin{equation*}
\int_{\Sigma_{U}} e^{-\beta H_{V}(\sigma)}\Pi(\ed(\sigma)) \geq
e^{m\beta(\# (D)-1)}\left(\frac{\lambda \pi
\varepsilon^2}{16}\right)^{\# (D)} e^{-\frac{\lambda \pi
\varepsilon^2}{16}\# (D)}
\end{equation*}
and it follows from \reff{fe} that
\begin{equation*}
P_{V}(\Omega^{1}(\mathcal{R})) \geq e^{m\beta(\#
(D)-1)}\left(\frac{\lambda \pi \varepsilon^2}{16}\right)^{\#
(D)}P_V(\Omega^0(\mathcal{R})).
\end{equation*}

Defining  $c(\beta)=e^{\beta m}$ and
 \begin{equation}\label{gfunction}
{G}(\beta,\lambda) = \beta m+\ln\lambda+\ln\left(\frac{\pi
\varepsilon^2}{16}\right)
\end{equation}
we obtain
\begin{align}\nonumber
P_V(\Omega^{0}(\mathcal{R}))&\leq\exp\{-\#
(D){G}(\beta,\lambda)\}\exp\{\beta m\}\\
\label{expest}
&=c(\beta)\exp\{-n(\mathcal{R})\alpha{G}(\beta,\lambda)\}.
\end{align}
The inequality \reff{mainlemma} holds in the infinite volume since
the right hand side of \reff{expest} does not depend on $V$.

Taking
$$
{\beta}_{\ell,h}^+(\lambda)=-\frac{1}{m}\ln(\lambda)-\frac{1}{m}
\ln\left(\frac{\pi \varepsilon^2}{16}\right)+\frac{h}{m}
$$
we complete the proof of Lemma \ref{l1}. \qed

\vspace{.5cm}

Next we define
\begin{equation} \label{b-plus}
\beta_{\ell}^+(\lambda)=\beta_{\ell,\frac{\ln(c)}{\alpha}}^+(\lambda),
\end{equation}
where $c$ is a combinatorial constant such that the number of the
contours of the length $n$ surrounding $0\in \R^2$ is not greater
than $c^n$. It is known that $c\leq 3$. Let $\lambda^+_\ell$ be
the solution of the equation $\beta^+_\ell(\lambda) = 0$. Define
the set
\begin{equation*}
A^+=\{(\beta,\lambda):\:\lambda\le \lambda^0,\
\beta>\beta^+_\ell(\lambda)\}   \cup \{ (\beta,\lambda):\:\lambda
> \lambda^+_\ell,\ \beta\geq 0 \}
\end{equation*}
\begin{lemma}\label{lest}
If $(\beta,\lambda)\in A^+$  then with the probability 1 there
exists only a finite number of $c$-contours  surrounding $0\in
\R^2$.
\end{lemma}
\pr Let $\Omega^{0}(\mathcal{R})$ be the set of all configurations
containing a $c$-contour $\mathcal{R}$, the empty contour  which
surrounds $0\in \R^2$, and let
$\Omega^{0}_n=\bigcup_{\mathcal{R}:\:n(\mathcal{R})=n}
\Omega^{0}(\mathcal{R})$. Then
$\Omega^{0}=\bigcup_{n}\Omega^{0}_{n}$ and
\begin{equation}\label{ine}
\sum_{n\geq 1}P^{\beta,\lambda}(\Omega^{0}_n)\leq \exp\{\beta
m\}\sum_{n\geq 1}\exp\left\{-n\Big(\alpha
G(\beta(\lambda),\lambda)-\ln(c)\Big)\right\}<\infty
\end{equation}
if $(\beta,\lambda)\in A^+$. It follows from \reff{ine} that
\begin{equation}\label{fin}
P^{\beta,\lambda}\left(\bigcap_{m}\bigcup_{n=m}^\infty\Omega^{0}_n\right)=0.
\end{equation}
The inequality \reff{fin} means that with the probability 1 there
exists a finite number of the empty contours surrounding 0. \qed

Let $\omega$ be a configuration. The set $Q_{\omega}=\bigcup_{x\in
\omega}B_{\ell / 2}(x)$ can be represented as the union of $\ell /
2$-neighborhoods of $\ell$-clusters which are  connected
components.

We define now a \textit{b-contour} (Boolean contour).  Assume that
there exists a line $L\subseteq Q^c_\omega$ surrounding $0\in\R^2$
such that $K_L\cap\omega=\varnothing$, where
\begin{equation*}
K_L=\bigcup_{x\in L}B_{\ell}(x).
\end{equation*}
The set $K_L$ is called a $b$-\textit{contour} surrounding 0 or
simply a $b$-\textit{contour}.

The $\ell/2$ neighborhood of any $\ell$-cluster does not intersect
$L$.

\begin{lemma}\label{curve} Assume that $\R^{2}$ is split
into cells of the linear size $q$. For  any b-contour $K$ with a
radius $r$, $r>\sqrt{2}q$, there exists a c-contour $\mathcal{R}$
such that $W(\mathcal{R})\subseteq K$.
\end{lemma}
\pr The proof is based on the following simple observation: if we
cast a coin of the radius $r$ on the plane $\mathbb R^2$ divided
into the square cells $\cal S,$ then there exists a cell which
will be covered entirely by the coin. Moreover if the center of
the coin lies on a boundary of two cells or four cells (one point)
then all those cells are covered by the coin. \qed

We say that two $b$-contours are different if $c$-contours
included into them are different. Since the number of the
$c$-contours is finite with the probability 1 then the number of
different $b$-contours is finite as well. Therefore there exists
an infinite component in $Q_\omega$ for almost all $\omega$. \qed

\section{Acknowledgements} The authors express their gratitude to
Pablo Ferrari for many valuable discussions. We thank Daniel
Takahashi for useful remarks. We thank Hans Zessin who pointed
M\"{u}rman's article out to us. We would like to express our
acknowledgements to both referees. They remarks stimulated the
improvement of the article results.

The work of E.P. was partly supported by Funda\c{c}\~{a}o de
Amparo \`{a} Pesquisa do Estado de S\~{a}o Paulo (FAPESP), the
grant $2008/53888-0$, and Russian Foundation for Basic Research
(RFFI) by the grants 08-01-00105, 07-01-92216.

The work of A.Ya. was partly supported by Conselho Nacional de
Desenvolvimento Cient\'{i}fico e Tecnol\'{o}gico (CNPq), the grant
306092/2007-7, and Programa de Apoio a N\'{u}cleos de
Excel\^{e}ncia (Pronex), the grant E26-170.008-2008.


\begin{thebibliography}{19}

\bibitem {Fernandez} Fern\'andez,~R.: Contour ensembles and the description of
  Gibbsian probability distributions at low temperature. Notes for a
  minicours given at the 21 Col\^oquio Brasileiro de Matem\'{a}tica, IMPA,
  Rio de Janeiro (1997)

\bibitem{KUH} Kac,~M., Uhlenbeck,~G and Hemmer,~P.C.: On the Van der
Waals Theory of Vapor-Liquid Equilibrium. J. Math. Phys., 5, 60-74
(1964)

\bibitem{LMP} Lebowitz,~J.L., Mazel,~A. and Presutti,~E.: Liquid-Vapor
Phase Transitions for Systems with Finite Range Interactions. J.
Stat. Phys., 94, 955-1027 (1999)

\bibitem {MM} Malyshev,~V.A., Minlos,~R.A.: Gibbs Random Fields, Cluster
Expansions. Kluwer Acad.Publ., Dordrecht (1991)

\bibitem{MS} Meester,~R. and Roy,~R.: Continuum Percolation. Series: Cambridge Tracts in Mathematics (No. 119). Cambrige
University Press (1996)

\bibitem{PZh} Pechersky,~E., Zhukov,~Yu.: Uniqueness of Gibbs State
for Non-Ideal Gas in ${\mathbb R}^d$: the case of pair potentials.
J. Stat. Phys. 97, 1/2, 145--172 (1999)

\bibitem{Ru} Ruelle,~D.:
Superstable Interactions in Classical Statistical Mechanics.
Commun. Math. Phys. 18, 127--159 (1970)

\bibitem{FPSY} P.~Ferrari, E.~Pechersky, V.~Sisko and A.~Yambartsev: Gibbs Random
Graphs,  in preparation.

\bibitem{S} Ya.G.~Sinai: Theory of phase transitions. Rigorous
results. Oxford, Pergamon Press, \vv{year}

\bibitem{Mur} Michael.G.~M\"{u}rmann.: Equilibrium Distribution of
Physical Clusters. Commun. Math. Phys. 45, 233--246 (1975).

\bibitem{zes} H. Zessin. A Theorem of Michael M\"{u}rmann revisited. J. Contemp.
Math. Anal. 43 (1), 68--80 (2008).

\bibitem{D}   R. L. Dobrushin. Gibbsian random fields for particles without hard
core.  Theoretical and Mathematical Physics, 4, 1, 705--719
(1970).

\end{thebibliography}
\end{document}